\documentclass[preprint2]{aastex}
\usepackage{graphicx}

\shortauthors{Hopkins et al.}
\shorttitle{Local Group SFH}

\begin{document}

\title{Global star formation history: A Local Group perspective}

\author{A. M. Hopkins\altaffilmark{1}(1), M. J. Irwin(2),
  A. J. Connolly(1)} 

\affil{\begin{enumerate}
\item Department of Physics and Astronomy, University of Pittsburgh, 
  3941 O'Hara Street, Pittsburgh, PA 15260, USA
\item Institute of Astronomy, Madingley Road, Cambridge, CB3 0HA, UK
\end{enumerate}
}

\altaffiltext{1}{Email: ahopkins@phyast.pitt.edu}

\begin{abstract}
The global star formation rate (SFR) density is estimated from the
star formation histories (SFHs) of Local Group galaxies. This is found to
be broadly consistent with estimates of the global SFH
from existing redshift surveys for two favoured cosmologies.
It also provides additional evidence for
a relatively constant global SFR density at high redshift ($z>1$).
\end{abstract}


\section{Introduction}
\label{int}

The global star formation history (SFH) of the universe has been probed
extensively by many different galaxy surveys, at many different wavelengths,
sampling regions from the local universe out to redshifts approaching $z=5$.
Despite uncertainties in star formation rate (SFR) calibrations,
the appropriate magnitude of extinction corrections at optical and
near-infrared wavelengths, and the heterogeneous nature of the many
contributing surveys, a remarkably consistent picture has developed.
From $z=0$ to $z=1$ it is generally agreed that $\rho^*$, the
comoving space density of the star formation rate (SFR) in galaxies,
rises by an order of magnitude \citep{Lil:96,Hog:98,Flo:99,Haa:00},
and between $1<z<2$ a flattening is seen \citep{Con:97,Yan:99,Haa:00,Hop:00}.
The behaviour of this evolution at higher redshifts however,
\citep{Mad:96,Ste:99,Hop:01}, is less well constrained.
It is still unclear whether
the evolution of $\rho^*$ reaches a peak around $z\approx1.5$ and decreases
significantly thereafter \citep[e.g.,][]{Mad:96,Gis:00}, or if it stays
flat to much higher redshifts \citep[e.g.,][]{Ste:99,Haa:00}.
One piece of evidence which may shed some light on this question
is the SFH of our Local Group of galaxies. Several recent analyses
\citep{PS:98,Fie:99,Tol:99,Rich:99,Gned:00,Hog:01,Avi:01} emphasise the
importance of comparing this ``fossil record" of our local Universe
to the data from high-redshift surveys, and we present such a
comparison in this Letter.

\section{Local Group SFH}

The properties of the Local Group have been the subject of several
recent reviews \citep{Mat:98,vdb:99,vdb:00}, and in particular the
SFH of many Local Group galaxies has been extensively explored using
various techniques \citep[e.g.,][]{Hod:89,Mat:98,Greb:00,Roch:00,Dol:00}.
These studies have achieved a level of sophistication sufficient to support
a comparison with the global SFH of the Universe from compilations
of redshift surveys. Such a compilation, incorporating suitable
extinction corrections, is shown in Figure~\ref{sfdnew},
taken from \citet[their Figure~6]{Hop:01}, for a
$(H_0,\Omega_{M},\Omega_{\Lambda})=(75,1.0,0.0)$ cosmology.
Due to the heterogeneous nature of the many surveys contributing to
this diagram and the varied methods for defining uncertainties by each,
a quantitative uncertainty estimate for the trend with redshift
is complex to define. Instead we simply consider an envelope (shown
in grey) encompassing the broadest of the published uncertainties,
and model predictions at higher redshifts. This will serve to
define the general trend, and a rough level of uncertainty, of
the observational results for comparison in later Figures.

The quantity being compared is $\rho^*$, the comoving space density of
the star formation rate, with units of $M_{\odot}\,$yr$^{-1}\,$Mpc$^{-3}$.
To construct this quantity for the Local Group we adopt a volume
defined by a 1.8\,Mpc radius sphere, ensuring that the faint
end of the Local Group luminosity function is consistent with
that of nearby field galaxies (Figure~\ref{lf}). This value is
consistent with the range from
1.2\,Mpc \citep{CV:99} to 1.8\,Mpc \citep{Mat:98} for the
radius of an idealised Local Group zero-velocity surface that
separates Hubble expansion from cluster contraction at the current epoch.

The SFHs of Local Group galaxies are typically presented in the
form of a relative SFR as a function of time and, to be
compared to an absolute measure, need to be normalised. We have
chosen to take current SFR estimates as the appropriate normalising
quantity. This is achieved by scaling the relative SFR values for each galaxy
to ensure the most recent time bin gives the current SFR. We adopt
$3\,M_{\odot}\,$yr$^{-1}$ for the current SFR of the Milky Way
\citep[e.g.,][]{Avi:01} and
$0.35\,M_{\odot}\,$yr$^{-1}$ for M31 \citep{Wal:94}.
The implications of this assumption are discussed in Section~\ref{disc} below.
The Milky Way (MW) is composed of many elements, the bulge, thin disk,
thick disk, halo, each with potentially different SFHs. Here we consider
only the contributions from the disk and bulge, the dominant
components of our Galaxy's SFH over its lifetime \citep{Gil:01}.
The SFH for the MW disk is taken from the recent results of \citet{Roch:00},
who emphasise its ``bursty" nature. No similar analysis appears
to be available for the MW bulge, a dominant component in the early
stages of the Galaxy's development, and to incorporate this contribution
we assume the bulge model of \citet[their Figure~3]{PS:98}. This model 
is already specified in absolute units, and thus does not require any
normalisation (the current SFR of the bulge from this model is
$\approx0.08\,M_{\odot}\,$yr$^{-1}$, a negligible contribution to the
current total). Similarly for M31, in the absence of detailed measurements
we have chosen to use the schematic form for the SFH of Sb galaxies
presented in \citet[Figure~8]{Ken:98}.
Clearly this type of smooth evolution is unlikely to be realistic, except 
perhaps in a coarse time-averaged sense, particularly as M31 is postulated
to have undergone one or more significant merger events over
the course of its evolution \citep{vdb:99}.
The SFHs for Local Group dwarf galaxies have been drawn from the review of
\citet{Mat:98}. Here, due to the lack of many reliable current SFR estimates,
we have normalised by the total masses of these objects, where available.
This leads to the inclusion of 25 Local Group dwarfs (of the 30 presented).
Since the total contribution from the dwarf systems is small anyway ($<10\%$
at all lookback times), the omission of five systems will not alter our
main conclusions significantly.

A ``global" SFH can now be constructed for the Local Group, and
is shown in Figures~\ref{sfh1} and \ref{sfh2}. These Figures
compare the Local Group measurements to the grey envelope of redshift
survey observations for two canonical cosmologies, (75,1.0,0.0) and
(75,0.3,0.7). In both Figures, the squares and stars represent the
SFH of the MW disk and bulge respectively, the triangles that of M31, and
the circles the contribution from the dwarf systems. The solid line is
the sum of these, representing the SFH of the Local Group. The
hashed region about this line indicates the effect of a factor
of 4 uncertainty in the current SFR of the MW and M31 (corresponding
to a range for the MW of $0.15-6\,M_{\odot}\,$yr$^{-1}$).

\section{Discussion}
\label{disc}

Given the extremely different methods used in measuring the variation
of $\rho^*$ with time for the Local Group and in redshift surveys,
it is highly encouraging to see the broad agreement shown in
Figures~\ref{sfh1} and \ref{sfh2}.
There are clearly discrepancies, too, which we explore here.
For both investigated cosmologies, there is a clear excess
in the Local Group $\rho^*$ in the most recent $\approx 3\,$Gyr. This
is attributable to two small Gyr scale ``mini-bursts" of
star-formation in the MW, from $0-1$\,Gyr and $2-3$\,Gyr. With
only two galaxies comprising the majority of the Local Group $\rho^*$
these small outbursts are given undue weight. To avoid this effect
a cosmologically representative volume would need to be sampled,
a factor of about 1000 greater than that investigated here, or
a $\gtrsim 30\,h^{-1}\,$Mpc cube \citep{Tur:92}. SFHs for at
least the most luminous (or massive) systems in this volume
would need to be determined.

The (75,1.0,0.0) cosmology of Figure~\ref{sfh1} immediately emphasises
the strong inconsistency in assumed ages for the Universe. This cosmology
implies an age of about 8.5\,Gyr, much less than the $>10\,$Gyr age
estimated for RR Lyrae stars and most Galactic Halo globular clusters,
and, particularly, the 14\,Gyr assumed in the modeling of the Local
Group SFHs. One result of this is a discrepancy between lookback
times of $5-7$\,Gyr. Although this is still possibly an artifact of small
number statistics, it does suggest that such a comparison might be
useful as yet another discriminator between various cosmological models.
This would require some refinement of the uncertainties in
both types of estimates for $\rho^*$, as well as modeling of the SFHs
for local systems with assumed ages corresponding to the cosmology being
tested.

One effect of choosing the current SFR for normalising the
relative SFHs is being able to compare an integrated ``mass of
stars formed" with measured masses and luminosities of the systems.
This results in $4.1\times10^{10}\,M_{\odot}$ for the MW (disk and bulge)
and $2.0\times10^{10}\,M_{\odot}$ for M31. While these values are
up to an order of magnitude smaller than estimates for the total
masses of these galaxies, they are roughly comparable to the total
{\em luminous\/} masses (with mass to light ratios of order unity in
solar units). The $V$-band luminosities of the MW and M31 are
$2.0\times10^{10}\,L_{\odot}$ and $2.6\times10^{10}\,L_{\odot}$
respectively, given $V_{\rm MW}=-20.9$ and $V_{\rm M31}=-21.2$
\citep{vdb:99} and using $V_{\odot}=+4.83$. This suggests that using total
masses instead of current SFRs for normalising the dwarf galaxy SFHs may
result in overestimating their contribution somewhat. This would then
have the effect of reducing the total Local Group SFH shown in
Figures~\ref{sfh1} and \ref{sfh2} by up to $\approx 5\%$.

\section{Conclusions}
\label{conc}

We have presented the SFH of the Local Group, expressed for the first
time as a global SFR density, $\rho^*$, and compared it with the results
from the many studies sampling this quantity as a function of redshift.
We find a broadly consistent magnitude for $\rho^*$,
within the expected uncertainties, for two favoured cosmologies.
Detailed discrepancies remain though, and directions
for future areas of research are suggested for addressing these.
The excess seen in the last 3\,Gyr can be attributed to 
two recent ``mini-bursts" of star formation in the MW, which have
an exaggerated effect due to the small number of galaxies
considered. To avoid this effect a suitably large volume of
local galaxies ($\gtrsim 30\,h^{-1}\,$Mpc cube) would need to be
analysed. At higher redshifts, the trend in $\rho^*$ from the Local Group
supports a fairly flat evolution. This is consistent with large
(factors of $\approx 10$) extinction corrections to high redshift
UV-based SFR measures.

\acknowledgements

AMH is grateful to Jeff Gardner and Alberto Conti for helpful discussion.
AMH and AJC acknowledge support provided by NASA through grant numbers
GO-07871.02-96A, LTSA NAG5-8546, and AR-07987.02-96A from the Space
Telescope Science Institute, which is operated by AURA, Inc., under
NASA contract NAS5-26555.

\begin{figure*}
\centerline{\rotatebox{-90}{\includegraphics[width=12cm]{sfdnew_col.ps}}}
\caption{Comoving SFR density ($\rho^*$) as a function of redshift. This
diagram is a compilation of SFR densities taken from \citet{Hop:01}. The
solid and dashed lines are models from \citet{Haa:00} and \citet{Gis:00}
respectively. The grey envelope has been defined to encompass these
measurements and indicate the general level of uncertainty. References are as
follows:
Haa00: \cite{Haa:00}; Hop00: \cite{Hop:00};
Ser00: \cite{Ser:00};
Sul00: \cite{Sul:00};
Gro00: \cite{Gro:00};
Ste99: \cite{Ste:99}; Yan99: \cite{Yan:99};
Tre98: \cite{Tre:98}; TM98: \cite{TM:98};
Con97: \cite{Con:97}; RR97: \cite{Row:97};
Gal95: \cite{Gal:95};
Con89: \cite{Con:89}.
 \label{sfdnew}}
\end{figure*}

\begin{figure*}
\centerline{\rotatebox{-90}{\includegraphics[width=12cm]{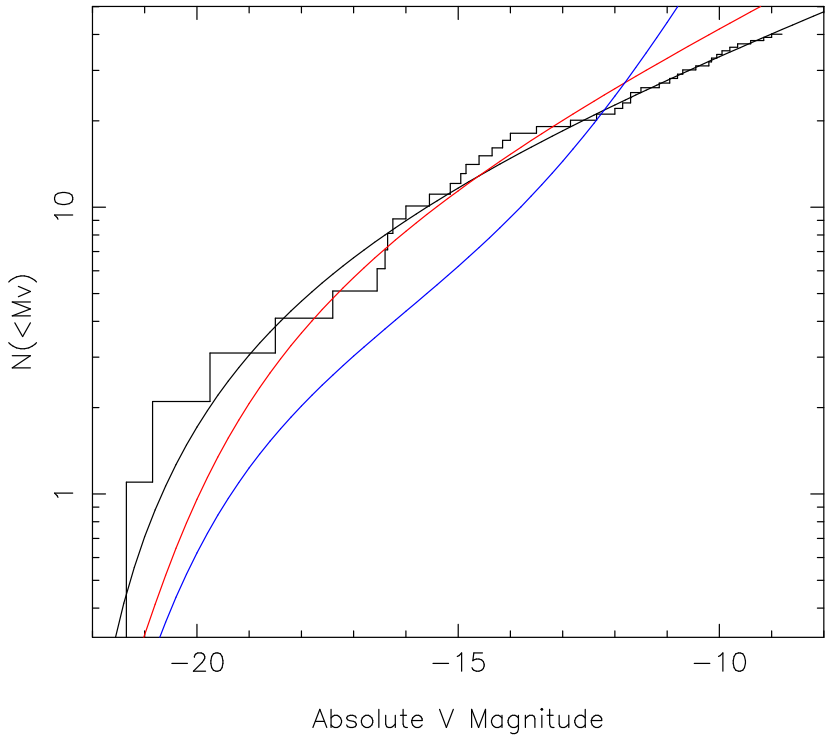}}}
\caption{Integral luminosity function for the Local Group galaxies
(histogram) using $H_0=75\,$km\,s$^{-1}$\,Mpc$^{-1}$,
and fitted Schechter function (solid black line). (Where necessary an
average value of $\left<B-V\right>=1$ was used for converting from $B$-band
to $V$-band magnitudes.)
The Schechter function parameters are $\phi^*=0.0081\,$Mpc$^{-3}\,$mag$^{-1}$,
$M^*=-22.6$ and $\alpha=-1.165$. The normalisation has been chosen
to match the faint ends of the LFs from \citet[dot-dashed (red) line]{Zuc:97}
and \citet[dashed (blue) line]{Mar:98} using a value of
$H_0=75\,$km\,s$^{-1}$\,Mpc$^{-1}$, and corresponds to a 1.8\,Mpc radius
sphere, consistent with other estimates of the zero-velocity surface for
the Local Group.
 \label{lf}}
\end{figure*}

\begin{figure*}
\centerline{\rotatebox{-90}{\includegraphics[width=9cm]{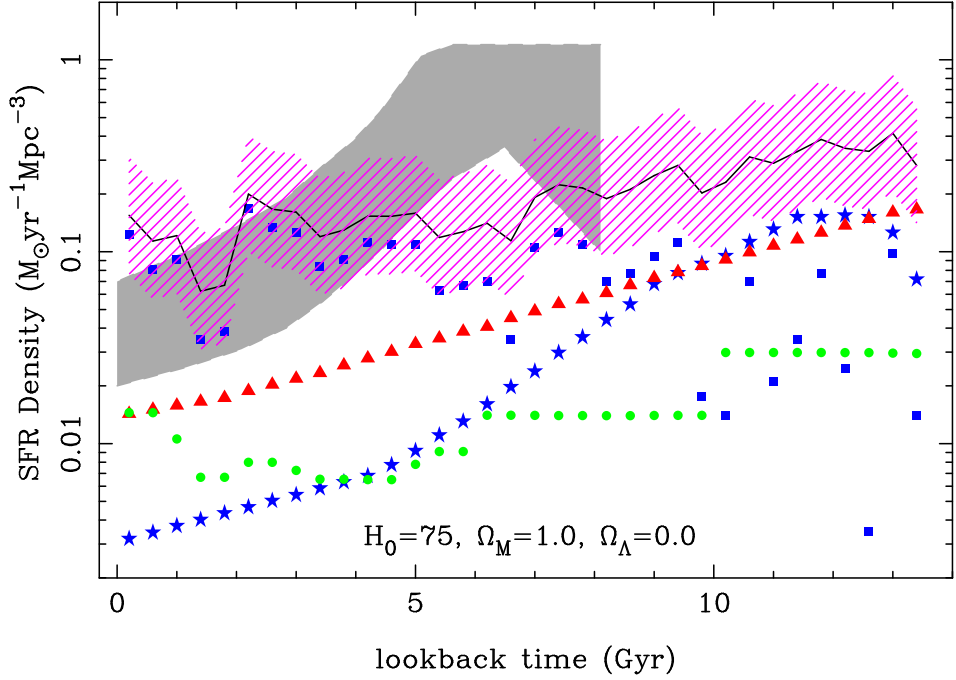}}}
\caption{Comparison of the Local Group SFH (solid line and hatched region)
with that from a compilation of redshift surveys (grey shaded area) for
a $(H_0,\Omega_M,\Omega_{\Lambda})=(75,1.0,0.0)$ cosmology. The squares,
stars, triangles and circles show the contributions from the MW disk,
MW bulge, M31 and dwarf galaxies respectively.
 \label{sfh1}}
\end{figure*}

\begin{figure*}
\centerline{\rotatebox{-90}{\includegraphics[width=9cm]{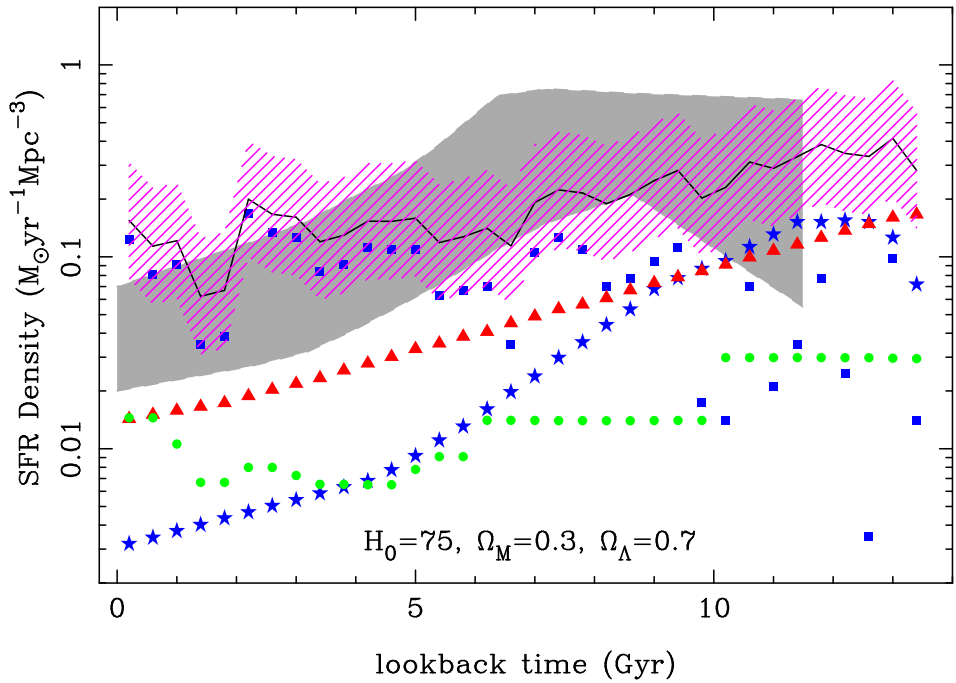}}}
\caption{As for Figure~\ref{sfh1} for a $(75,0.3,0.7)$ cosmology.
 \label{sfh2}}
\end{figure*}

\end{document}